\documentclass[%
 reprint,
 superscriptaddress,
 amsmath,amssymb,
 aps,
 prb
]{revtex4-2}

\usepackage{braket}
\usepackage{graphicx}
\usepackage{dcolumn}
\usepackage{bm}
\usepackage{siunitx}
\usepackage{xcolor} 
\usepackage{tabularx}
\usepackage{dsfont}					
\usepackage{mathrsfs}				
\usepackage{float}
\usepackage{hyperref}

\newcommand{\be}[0]{\begin{equation}}	
\newcommand{\ee}[0]{\end{equation}}

\usepackage{hyperref}
\hypersetup{colorlinks, 
	linkcolor={blue!75!black!80!yellow},
	citecolor={blue!75!black!80!yellow}, 
	urlcolor={blue!75!black!80!yellow}
	}

\usepackage[normalem]{ulem}

\makeatletter
\DeclareRobustCommand{\equalcontrib}{%
  \frontmatter@footnote{These authors contributed equally to this work.}%
}
\makeatother

\begin{document}


\title{Stabilizing fluctuating Cooper pairs with a terahertz resonator}

\author{Frieder Lindel\equalcontrib}
\email{flindel@phys.ethz.ch}
\affiliation{Institute for Theoretical Physics, ETH Zürich, Zürich 8093, Switzerland}
\affiliation{Quantum Center, ETH Zürich, Zürich 8093, Switzerland}

\author{Jonathan B. Curtis\equalcontrib}
\email{jon.curtis.94@gmail.com}
\affiliation{Institute for Theoretical Physics, ETH Zürich, Zürich 8093, Switzerland}

\author{Jérôme Faist}
\affiliation{Quantum Center, ETH Zürich, Zürich 8093, Switzerland}
\affiliation{Institute of Quantum Electronics, ETH Zürich, Zürich 8093, Switzerland}

\author{Eugene Demler}
\email{demlere@phys.ethz.ch}
\affiliation{Institute for Theoretical Physics, ETH Zürich, Zürich 8093, Switzerland}

\date{\today}

\begin{abstract}
Coupling quantum materials to electromagnetic cavities or resonators without external driving has gained prominence as a new control mechanism. Here we develop an equilibrium theory of a mesoscopic, disordered two-dimensional superconductor coupled to a subwavelength terahertz resonator, focusing on the interplay between superconducting phase fluctuations in the sample and electromagnetic fluctuations of the resonator. Modeling the resonator as a lumped-element circuit, we show that the coupling enhances, rather than suppresses, the superconducting transition temperature $T_c$. When the sample is smaller than the Pearl length, phase fluctuations near the Berezinskii–Kosterlitz–Thouless (BKT) transition strongly amplify this enhancement. Crucially, the resistive quasiparticle response of the sample must be included in the analysis: at low frequencies the superfluid shunts the normal-fluid channel, reducing the total sample impedance. For representative parameters, a $1 \unit{\micro\meter}  \times 1\unit{\micro\meter}$, $8 $nm-thick disordered film, we find a relative $T_c$ enhancement of about $1\%$, an order of magnitude larger than the $0.1\% $ obtained when phase fluctuations are neglected. These results identify superconductors with soft phase fluctuations as promising platforms for equilibrium control with terahertz resonators.
\end{abstract}

\maketitle

\begin{figure}
    \centering
    \includegraphics[width=\linewidth]{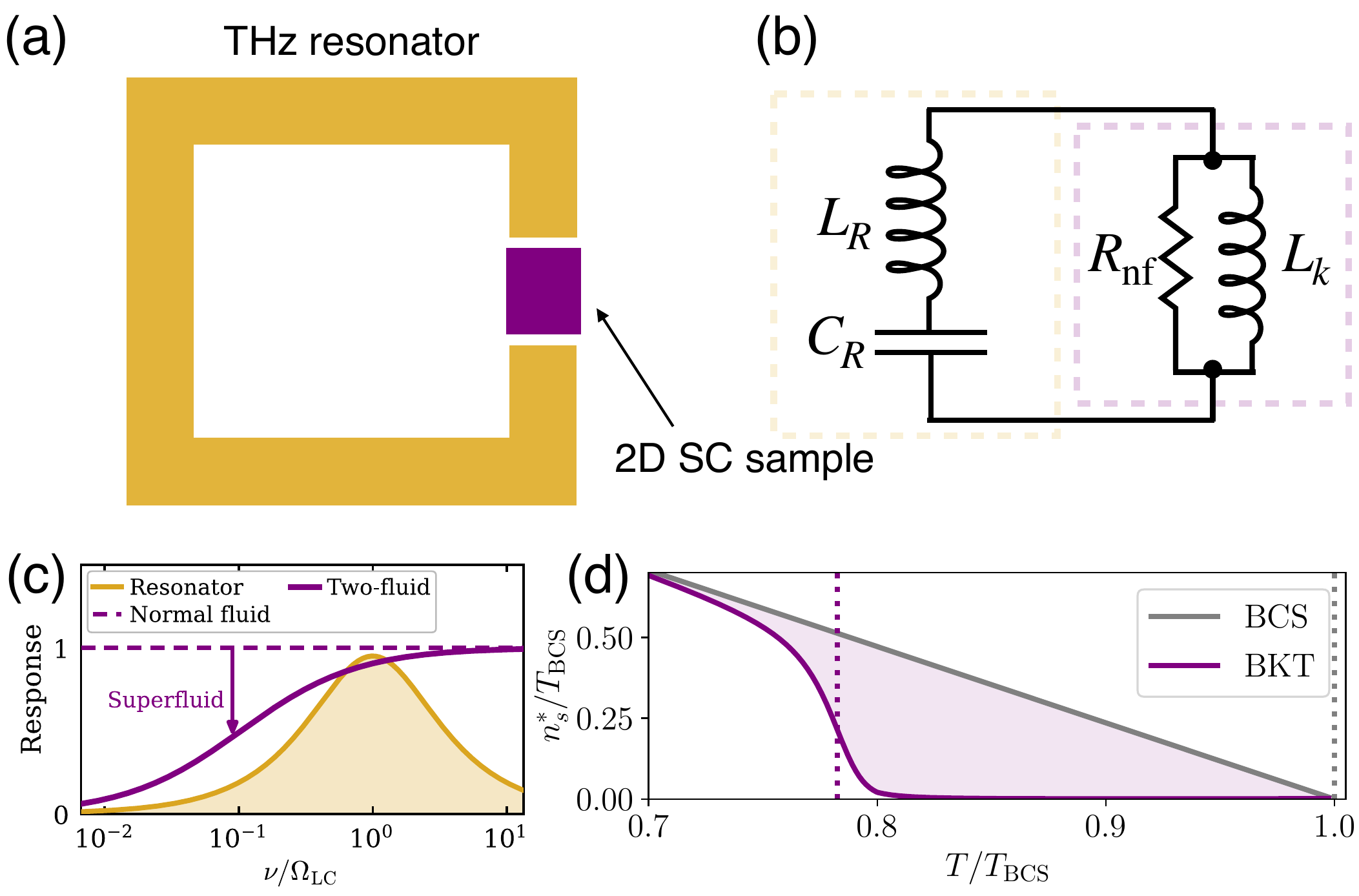}
    \caption{(a) Illustration of a terahertz (split-ring) resonator interacting with a two-dimensional superconducting (SC) sample.
    (b) Effective circuit description of the resonator and sample within the two-fluid model approximation.
    (c) Basic mechanism of enhancement. 
    The impedance of the sample (purple line) drops at low frequencies in the superconducting state \emph{in comparison to the normal state} (dashed line) due to the superfluid condensate shunting the dissipative normal fluid. The relevant resonator free energy is proportional to the overlap of the sample impedance with the resonator spectrum ($1/Z_\mathrm{R}$, yellow line), the precise mathematical statement is given in Eq.~\eqref{eqn:actioncircuit} (see also Appendix~\ref{app:circuitQED}), and the system seeks to minimize this. (d) Superfluid density $n_s^\ast$ as a function of temperature within the bare BCS framework and accounting for fluctuations realizing a Berezinskii-Kosterlitz-Thouless (BKT) transition. Dotted lines mark $T_c$, defined as the onset of $n_s^\ast$ (BCS) and its steepest drop (BKT).}
    \label{fig:schematic}
\end{figure}

{\it Introduction.}\textemdash 
The optical control of quantum materials typically relies on illuminating the materials with classical radiation with unwanted side effects such as heating. Recently, a novel route towards optical control of quantum materials has been demonstrated. Here, the sample is coupled to a cavity or resonator, which modifies the \emph{equilibrium} thermal and quantum fluctuations of the electromagnetic field and thereby also the equilibrium properties of the sample \cite{schlawin2022cavity,bretscher2026fluctuation,lu2025cavity}. This has been used to e.g.~modify topologically protected quantum Hall states \cite{enkner2025tunable,appugliese2022breakdown}, to align quantum Hall stripes \cite{graziotto2026cavity}, to demonstrate cavity-induced modifications of the superfluid density and changes in the critical temperature of superconductors of a few percent \cite{wang2026evidence,keren2026cavity,xu2026vacuum,zhang2026cavity,montanaro2026cavity}, and, via cavity-modified thermal exchange, to switch a metal-to-insulator transition~\cite{jarc2023cavity}.

Existing theories of cavity-modified superconductivity mostly consider cavity-induced modifications of a microscopic coupling constant, such as the pairing interaction \cite{schlawin2019cavity,plastovets2025cavity,riolo2025polariton}, the electron-phonon interaction \cite{sentef2018cavity,lu2024cavity}, or suppressed electron hopping \cite{kozin2025cavity}, or consider nonequilibrium effects~\cite{Curtis.2019}. Alternatively, a perturbative expansion of the minimally coupled Ginzburg--Landau functional has been analyzed in Refs.~\cite{wang2026evidence,jiang2026no}. The percent-level shifts of $T_{\rm c}$ and changes in the superfluid density or optical response seen in experiments \cite{wang2026evidence,keren2026cavity,xu2026vacuum,zhang2026cavity} remain quantitatively unexplained.

Recently, the idea of Casimir control has been developed \cite{carlsson2026casimir,curtis2023local,kotov2025casimir}: the equilibrium ground-state (Casimir) energy of the electromagnetic field inside the cavity is computed as a function of the material's optical properties, which themselves depend on the microscopic state of the sample. The Casimir energy then enters
the free-energy landscape of the order parameter as a `vacuum pressure' exerted by the field fluctuations. For superconductors this has a longer history, but typically asked the other way round: the Casimir free energy of a superconductor in close proximity to a metal plate has been studied in order to probe the changes in the Casimir energy itself across the superconducting phase transition \cite{villarreal2019casimir,bimonte2005towards,bimonte2005variations,bimonte2006casimir,bimonte2025proposal}. Seeing these effects in the single-metal-plate geometry has so far remained elusive in experiments \cite{perez2020system,campbell2025seeking}, in contrast to the recent experiments in which the samples were embedded in sub-wavelength resonators instead \cite{wang2026evidence,zhang2026cavity}.

Here, inspired by these recent experiments~\cite{wang2026evidence,zhang2026cavity}, we study how the critical temperature $T_{\rm c}$ of a dirty two-dimensional superconductor is modified when the sample is placed inside a sub-wavelength THz resonator; see Fig.~\ref{fig:schematic}. Following the Casimir control framework, our starting point is the equilibrium free energy of the electromagnetic fluctuations of the resonator mode modeled as a lumped circuit \cite{barash2026casimir,shahmoon2018electronic,%
shahmoon2017casimir}, which depends on the properties of the superconducting sample. Adding this contribution to the Ginzburg--Landau functional yields the cavity-induced changes of the free-energy landscape of the order parameter, from which we extract modifications in the critical temperature $T_{\rm c}$, defined as illustrated in Fig.~\ref{fig:schematic}(d) \footnote{The definitions of the critical temperature for a finite-sized two-dimensional sample in the bare BCS framework and in the presence of fluctuations are also given below Eqs.~\eqref{eq:nstae_BCS} and \eqref{eq:nstar_BKT}, respectively.}. For a dirty $1 \unit{\micro\meter}  \times 1\unit{\micro\meter}$, $8 $nm-thick superconductor with low normal state sheet conductivity $\sigma_\square =0.64 e^2/\hbar $ and zero temperature coherence length $\xi_\mathrm{GL}(0) = 10\,\mathrm{nm}$, we find a resonator-induced enhancement of $T_{\rm c}$ of the order $\gtrsim 1\%$ compared to a sample of the same size but without the cavity. In contrast to previous minimally coupled Ginzburg-Landau treatments \cite{wang2026evidence,jiang2026no}, this approach is non-perturbative in the light-matter coupling; it relies on taking the resistive normal state background of the superconductor into account, which changes the predicted cavity-induced decrease in $T_c$ into a $T_c$ enhancement; it accounts for phase fluctuations of the order parameter described by Berezinskii-Kosterlitz-Thouless (BKT) physics, without which, we find that the relative shift in $T_{\rm c}$ is limited to $0.1\,\%$, showing that the softening of the transition due to fluctuations is crucial. While the dirty-limit disordered superconductor considered here differs from the material of Ref.~\cite{wang2026evidence}, we argue that a similar mechanism may be at play.

{\it Circuit model.}\textemdash  We start by describing the split ring resonator under consideration, which consists of a metallic structure [Fig.~\ref{fig:schematic}(a)] that can be modeled as a lumped $LC$ circuit. Its metal arms provide the inductance $L_\mathrm{R}$ and the small gap between the arms its capacitance $C_\mathrm{R}$. This leads to the resonance frequency of the bare split ring resonator without the sample $\Omega_{\rm LC} = 1/\sqrt{C_{\rm R}L_{\rm R}}$, which typically resides in the THz frequency range, a ratio $(C_R/e^2)/(e^2 L_R)$ [setting $\hbar = 1$] which typically is on the order of $1-800$ \cite{jeannin2019ultrastrong,valmorra_PhD}, and a frequency dependent resonator impedance $Z_\mathrm{R} =\mathrm{i} [ 1/(\omega C_\mathrm{R}) -\omega L_\mathrm{R}]$. The equilibrium free energy describing its quantum and thermal fluctuations is that of a single harmonic oscillator mode \cite{barash2026casimir} (see Appendix~\ref{app:circuitQED}).

When the sample is coupled to the split-ring resonator, this shifts the total impedance of the circuit and consequently its fluctuations and free energy. We consider that the superconducting sample with impedance $Z_\mathrm{sc}$ is placed in series with the split-ring resonator, so that the total impedance of the circuit with the sample reads $Z= Z_\mathrm{R} + Z_\mathrm{sc}$. This can be achieved by galvanically coupling the sample to the resonator or by placing it in the capacitive region similar to Ref.~\cite{wang2026evidence}, see Appendix \ref{app:capacitors}. Appealing to the Caldeira-Leggett~\cite{Caldeira.1983rkp} prescription for dissipation in quantum systems we find that the Euclidean effective action of the coupled circuit, up to terms that are independent of the superconducting properties and the bare sample action, is given by (see Appendix~\ref{app:circuitQED})
\begin{equation}
    \label{eqn:actioncircuit}
    {S}_{\rm LC}[n_s] = \frac12 \sum_{\nu_m} \log\left[ 1 + \frac{Z_\mathrm{sc}(\mathrm{i}\nu_m)}{Z_R(i\nu_m)}\right].
\end{equation}
Here, the sum runs over bosonic Matsubara frequencies $\nu_m = 2\pi mT$ and the Euclidean action equals the free energy $F$ in units of temperature $T$, $S = F/T$.

{\it BCS theory.}\textemdash We now turn our attention to describing the superconducting sample.
We will assume for simplicity that we can describe our sample by a dirty BCS superconductor with the action as a functional of the unitless superfluid density $K = n_s/T$ near the critical temperature without the resonator $T_\mathrm{BCS}$ (such that we can expand in powers of $K$) of (setting $\hbar = k_\mathrm{B}   = 1$)
\begin{equation}
    \label{eqn:actionBCS}
    {S}_{0}[K] = \frac{1}{2} \left[  \textrm{sgn}(t-1) K   + \frac{7 \zeta(3) e^2}{2 \pi^3  \sigma_\square } \frac{t}{|1-t|}K^2  \right] .
\end{equation}
Here we have given the action for a patch of area $\xi_{\rm GL}^2(t)$, expressed it in terms of the reduced temperature $t = T/T_{\rm BCS}$, and we have used the dirty superconductor results for coherence length $\xi_{\rm GL}^2(t) = \pi D/[8 T_\mathrm{BCS}(1-t)]$ and two-dimensional superfluid density $n_s = \pi \sigma_\square \Delta^2/(8 e^2 T_\mathrm{BCS}) $, where $\Delta$ is the superconducting gap, $\sigma_\square$ the normal state sheet conductivity, and $D$ the diffusion constant.
Crucially, for a sample of size $L^2$ the BCS action scales extensively to yield the total action of $(L/\xi_{\rm GL})^2
S_0[n_s]$.
Note that here $n_s$ is a free degree of freedom, corresponding to the amplitude of the superconducting order parameter; the standard BCS formula for $\Delta(T)$ can then be recovered by extremizing~\eqref{eqn:actionBCS} over $n_s$. In this case, the central free parameter is $\sigma_\square = 2e^2\nu_F d D$, with $\nu_F$ the electronic density of states per spin. We assume that the film thickness $d$ is smaller than $\xi_\mathrm{GL}(0)$ such that the order parameter is uniform across the film thickness.

The terahertz resonator interacts through the effective circuit depicted in Fig.~\ref{fig:schematic}(b), which means we need the impedance of the sample as a functional of the superfluid density $n_s$, evaluated at Matsubara frequencies; compare Eq.~\eqref{eqn:actioncircuit}.
In the two-fluid approximation, we find the impedance per square of the sample 
\begin{equation} \label{eq:Zsquar}
    Z_{\mathrm{sc}\square}[i\nu_m; n_s] = \left[ \frac{1}{R_{\mathrm{nf},\square}}\textrm{sgn}(\nu_m) + \frac{4e^2n_s}{ \nu_m}\right]^{-1},
\end{equation}
where $Z_\mathrm{sc} = N_{\square} Z_{\mathrm{sc}\square}$, with the number of squares $N_{\square}$, and the factor of $4e^2$ reflects the charge $2e$ of Cooper pairs relative to their phase stiffness. $Z_\mathrm{sc}$ is therefore given by the normal-state resistance per square $R_{\mathrm{nf},\square} = 1/\sigma_\square$ in parallel with the superconductors kinetic inductance per square $L_{k,\square} = 1/(4 e^2 n_s)$ of the sample; see Fig.~\ref{fig:schematic}(b). Using Eq.~\eqref{eq:Zsquar} to evaluate the free energy in Eq.~\eqref{eqn:actioncircuit}, we find that in the presence of Ohmic dissipation, this is logarithmically UV divergent, but can be renormalized by referencing to the normal state of the sample.
Ultimately, the sum will converge once the inductance of the normal state is also resolved, as is captured, for instance, in the Zimmermann model \cite{zimmermann1991optical}; see Appendix~\ref{app:bare_BCS_zimmermann}.

\begin{figure}
    \centering
    \includegraphics[width=\linewidth]
    {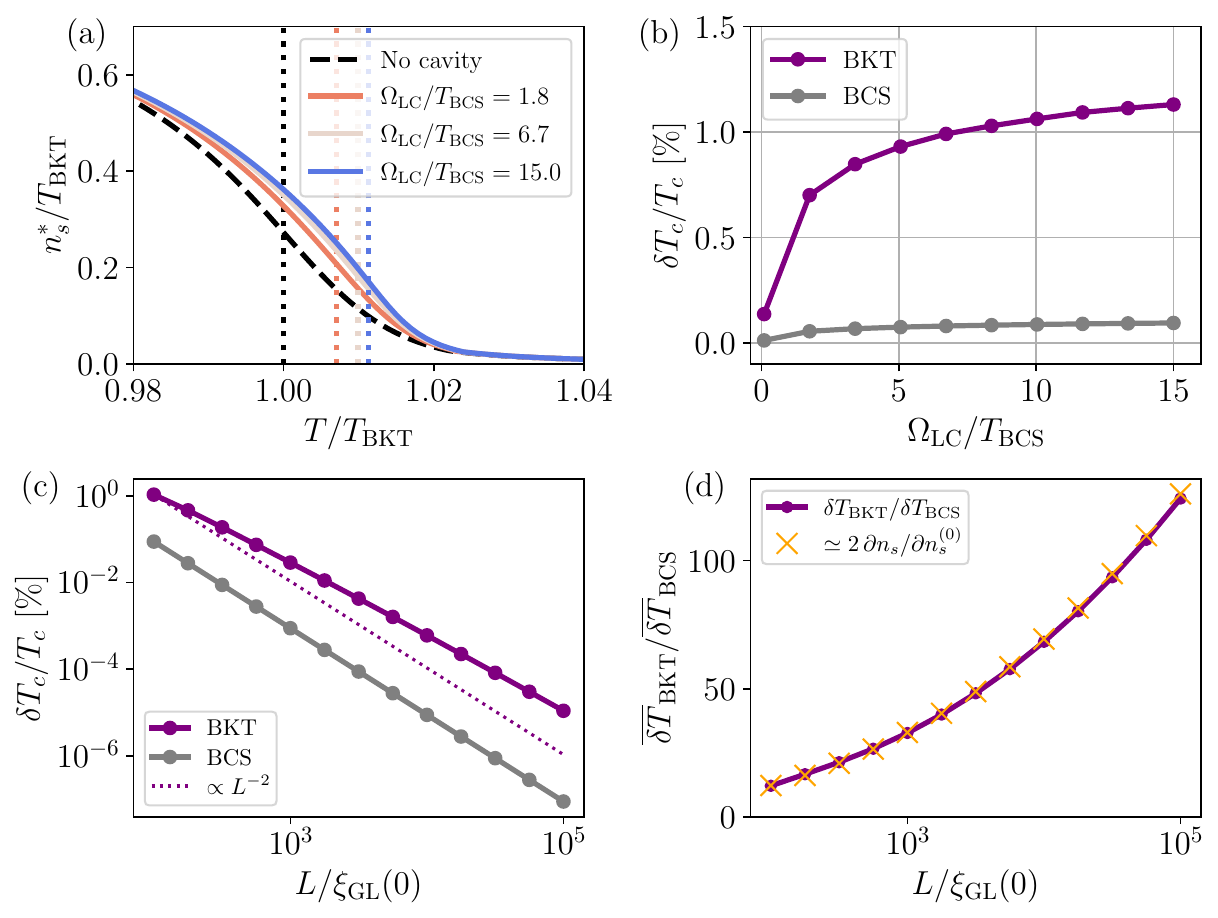} 
    \caption{(a) Renormalized superfluid density $n_s^\ast$ as a function of temperature $T$ without the cavity (black dashed line) and with cavity of increasing frequency $\Omega_\mathrm{LC}$ using $\sigma_\square = 0.64 e^2/\hbar$, $L/\xi_{\rm GL}(0) = 100$, and $(C_R/e^2)/(e^2 L_R) = 400 $. The vertical dotted lines indicate the critical temperature for each configuration defined as the point of steepest descent of $n_s^\ast(T)$. (b) Relative shift of the critical temperature $\delta T_c$ (purple line) compared to the critical temperature $T_c$ of the same sample without the cavity. $\delta T_c$ increases with increasing resonator frequency, reaching $ 1\%$ at $\Omega_{\rm LC} \sim 6.7 T_{\rm BCS}$, while remaining below $0.1\%$ neglecting fluctuation effects (bare BCS, gray line). (c) Relative shift of the critical temperature as a function of sample size $L/\xi_{\rm GL}(0)$ for $\Omega_{\rm LC} = 10\,T_{\rm BCS}$, for the BKT (purple) and bare BCS (gray) models. The dotted line indicates $\propto L^{-2}$. (d) Ratio of the BKT and bare BCS critical temperature shifts (dots) compared with the estimate of Eq.~\eqref{eq:ratio} (crosses).
    }
    \label{fig:BKT}
\end{figure}

{\it Bare BCS control.}\textemdash We now minimize the action over the superfluid density to find the effect of the resonator on $T_\mathrm{BCS}$.
First, we obtain 
\begin{equation} \label{eq:nstae_BCS}
    n_s^*(T) = \mathrm{arg} \min_{n_s\geq 0}\left\{ S_{\rm LC}[n_s] + \left(\frac{L}{\xi_{\rm GL}}\right)^2 S_0[n_s]\right\}.
\end{equation}
We define the critical temperature as the largest temperature with nonzero $n_s^*$ and denote its resonator-induced shift by $\delta T_\mathrm{BCS}$. For typical split ring resonators with $(C_R/e^2)/(e^2 L_R) = 400$ \cite{jeannin2019ultrastrong} and taking $\sigma_\square = 0.64$ and $L/\xi_{\rm GL}(0) = 100$ we find an increase in the critical temperature of about $0.1\%$ at $\Omega_\mathrm{LC} = 15 T_\mathrm{BCS}$, see the gray line in Fig.~\ref{fig:BKT} (b) and Appendix~\ref{app:bare_BCS_zimmermann} for details. For an uncoupled critical temperature $T_{\rm BCS} \sim 6.4$K this corresponds to a resonator frequency $\Omega_{\rm LC}/2\pi \sim 2 $THz similar to Ref.~\cite{wang2026evidence}. Further setting $\xi_\mathrm{GL}(0) = 10\,$nm, which is comparable to that in NbSe${}_2$ \cite{nader2014critical} or disordered thin films \cite{yadav2023probing}, we have $L = 1\,\mu \mathrm{m}$. For a film thickness $d = 8$\,nm, these parameters correspond to $D \approx 2\,$cm$^2$/s and $\nu_F \approx 2\times10^{45}\,$J$^{-1}$m$^{-3}$, as realized in strongly disordered films with low carrier density \cite{shahar1992superconductivity,sacepe2020quantum}. Without fluctuation effects considered in the next section, we therefore predict that a (small) enhancement in the transition temperature should be observed when such a sample is placed in the resonator.

In order to understand the origin of the enhancement of $T_\mathrm{BCS}$, it is enough to consider the vacuum energy of the resonator, which, in the absence of dissipation and thermal fluctuations, is $\hbar \Omega_\mathrm{LC}/2 = \frac{\hbar}{2\sqrt{L_R C_R}}$. 
When coupled to the normal metal, the impedance of the resonator is dramatically affected by the normal state resistance, which typically damps the resonator and reduces the energy gain in the resonator due to vacuum fluctuations. Upon cooling below $T_\mathrm{BCS}$ the condensate shunts the resistance and in general reduces the impedance of the sample at the relevant frequencies; compare Fig.~\ref{fig:schematic} (c).
This then improves the resonator, as has been observed experimentally in Ref.~\cite{scalari2014high}, and increases the energy gain from zero-point fluctuations. It is therefore important to include the normal-fluid component of the sample which is present at finite frequencies and above $T_\mathrm{BCS}$; the inclusion of the normal-fluid response has often been overlooked in recent theoretical works \cite{wang2026evidence,jiang2026no,plastovets2025cavity}, but is a crucial driver of the enhancement~\footnote{Neglecting the normal fluid impedance would lead to an increase in the total inductance upon decreasing $n_s$ such that the resonator would actually favor the normal state}\cite{bimonte2005towards,bimonte2010optical}. 
We also see that the effect is expected to be small for a realistic sample size due to the circuit action $S_{\rm LC} \sim O(1)$ being single-mode, whereas the sample condensate action is $(L/\xi_\mathrm{GL})^2 S_0 \sim O(L^2)$. Equivalently, a single resonator mode competes with the condensation energy of all electrons participating in pairing, whose number $N \sim (\sigma_\square\hbar/e^2)(L/\xi_{\rm GL}(0))^2$ is of order $10^3$ for our parameters, and $\delta T_c$ decreases roughly as $1/N$. An appreciable effect therefore requires small, strongly disordered samples with few participating electrons. This is in contrast to recent proposals where multimode Casimir forces were shown to provide a route towards controlling nematic phases in electronic materials \cite{carlsson2026casimir}. 
We now show how, by including the strong fluctuation effects present in two dimensions, the effect of the resonator can be greatly enhanced.

{\it Fluctuation effects.}\textemdash We have seen that in principle a terahertz resonator can enhance the transition temperature of a superconducting sample due to the free energy gain in the superconducting state, although typically the effect is small, scaling with system size as $\delta T_\mathrm{BCS}/T_\mathrm{BCS}\sim (\xi_{\rm GL}/L)^2$ and therefore for a micron-scale sample amounts to a relative shift typically of order $10^{-3}$.  

This however points to an important case where the effect may be substantially larger\textemdash in systems with strongly fluctuating superconductivity. 
As a conceptually clean and practically relevant case, let us consider a two-dimensional superconducting thin film with strong disorder, whose size is below the Pearl length, i.e. $L < 2 \lambda_L^2/d$  with
$\lambda_L$ the London penetration depth, such that vortices interact logarithmically across the entire sample. In this case, it is known that the superconducting transition can be significantly lowered and broadened due to vortex-antivortex fluctuations which realize a Berezinskii--Kosterlitz--Thouless (BKT) transition \cite{Kosterlitz.1974,Kosterlitz.2016,Beasley.1979}.

In order to capture these effects, we employ the renormalization group (RG) approach outlined by Kosterlitz~\cite{Kosterlitz.1974,Kosterlitz.2016}, with scaling equations for superfluid stiffness $K = n_s/T$ and vortex fugacity $y$ which read 
\begin{subequations}
\begin{align}
    & \frac{d K}{db} = -4\pi^3 y^2 K^2 ,\\
    & \frac{dy}{db} = (2-\pi K)y ,
\end{align}
\end{subequations}
with $b = \log \ell /\xi_{\rm GL}$ the RG scale parameter.
In particular, the relevant degree of freedom with which the circuit interacts is the superfluid density at the sample scale $b = \log L/\xi_{\rm GL}$, which includes the strong renormalization due to vortex-antivortex screening. 
Given the bare dimensionless stiffness $K_0 = n_s^{(0)}/T$, vortex fugacity $y_0 \propto e^{-\pi K_0/2}$, and action $S_0$ of a $\xi_\mathrm{GL}^2$ patch from Eq.~\eqref{eqn:actionBCS}, we can integrate these equations to find $K^R = n_s /T$.
We then express the BCS action $S_0[n_s(n_s^{(0)})]$ as a functional of the renormalized superfluid density. For a discussion of whether the bare or renormalized action should be used, see Appendix~\ref{app:BKT}. Also note that here we use a simple two-fluid model to describe the ac response of the sample as a functional of the superfluid density; incorporating the full dynamically renormalized response~\cite{Ambegaokar.1980,Halperin.1979} is left for future investigations.

Minimizing over the renormalized superfluid density $n_s$ we find 
\begin{equation} \label{eq:nstar_BKT}
    n_s^*(t) =\mathrm{arg} \min_{n_s \geq 0}\left\{ S_{\rm LC}[n_s]+ (L/\xi_{\rm GL})^2 S_0[n_s^{(0)}(n_s)] \right\}.
\end{equation}
For a sample of finite size $L$, $n_s^*(t)$ drops continuously [compare Fig.~\ref{fig:schematic}(d)] over a temperature window that narrows as $L$ grows \cite{Nelson.1977}. We define the critical temperature $T_\mathrm{c}$ of the finite sample as the temperature at which $n_s^*(T)$ drops most steeply, in which case $T_c$ approaches the thermodynamic BKT temperature for $L\to \infty$. Without the resonator, we denote this temperature by $T_{\rm BKT}$.

 The result is shown in Fig.~\ref{fig:BKT} using $(C_R/e^2)/(e^2 L_R) = 400$, $\sigma_\square = 0.64$ and $L/\xi_{\rm GL}(0) = 100$, as before. We see in Fig.~\ref{fig:BKT}(a) that the renormalized superfluid density exhibits a sharp drop near $T_\mathrm{BKT}$ due to vortex-antivortex fluctuations. Moreover, coupling the resonator to the sample leads to a notable increase in the transition temperature [vertical dotted lines in Fig.~\ref{fig:BKT}(a)] and to an increase in the renormalized superfluid density. Increasing the resonator frequency (at fixed characteristic resonator impedance $\sqrt{L_R e^4/C_R}$) increases the enhancement of $T_\mathrm{BKT}$, reaching scales $\gtrsim 1 \%$ for a resonator frequency of $\Omega_\mathrm{LC} = 6.7T_\mathrm{BCS}$, as shown in Fig.~\ref{fig:BKT}(b). If we estimate the uncoupled critical temperature $T_{\mathrm{BKT}} \sim 6.4$K then this corresponds to a resonator frequency of $1.1$\,THz, comparable to the resonators of Ref.~\cite{wang2026evidence}.

This increase in the cavity-induced effects compared to the bare BCS model can be attributed to the sample's sensitivity near the transition temperature $T_\mathrm{BKT}$: as $L$ grows larger, the drop in $n_s^\ast$ near $T_\mathrm{BKT}$ becomes sharper and sharper~\cite{Nelson.1977}, such that only a small change in the bare superfluid density $n_s^{(0)}$ is needed to effect a large change in the circuit impedance, albeit over an increasingly narrow temperature regime as $L\to \infty$. Consequently, while the sample condensation energy scales extensively with $L$, only a very small change in this energy is needed to balance the change in the circuit's zero-point energy. To quantify this effect, we find to linear order in $S_{\rm LC}$
\begin{equation} \label{eq:ratio}
 \frac{\overline{\delta T}_{\mathrm{BKT}}}{\overline{\delta T}_{\mathrm{BCS}}} \simeq
  2\,\frac{\partial n_s}{\partial n_s^{(0)}}\bigg|_{T_{\mathrm{BKT}}},
\end{equation}
where $\overline{\delta T}_c = \delta T_c/T_c$.
As the drop in $n_s$ sharpens with $L$, the enhancement over the bare BCS shift increases with $L$, compare Fig.~\ref{fig:BKT}(d). However, this does not compensate for the $(\xi_{\rm GL}/L)^2$
suppression: $\delta T_\mathrm{BKT}$ still decreases with $L$, albeit more slowly than $\delta T_\mathrm{BCS} \propto L^{-2}$.

While our theory correctly predicts both the sign (positive) and rough order of magnitude ($\sim1\%$) scale of the $T_c$ enhancement observed experimentally \cite{wang2026evidence}, our theory is not consistent with the normal-state resistance of the samples used in these experiments \cite{wang2026evidence,zhang2026cavity}.
In our models, we find that BKT physics is only appreciable at these scales for substantially lower normal state conductivity, $\sigma_\square \sim 0.64$, in contrast to recent experiments \cite{wang2026evidence} which saw similar BKT physics emerge for $\sigma_\square \sim 160$ \textemdash orders of magnitude cleaner. 
It is then extremely surprising that strong fluctuation effects were observed with such a large normal state conductivity, perhaps hinting that BKT physics in this system is driven by a different mechanism than the standard culprit of strong disorder~\cite{Beasley.1979}.
In this light, it is interesting to note that the material NbSe$_2$ also has a strong charge-density-wave order which may compete with superconductivity and could lead to enhanced phase fluctuations even in a clean sample. Also, the in-plane optical scattering rate of NbSe$_2$ increases with
frequency~\cite{dordevic2001anisotropic}, so that at the terahertz Matsubara frequencies relevant
here, the normal fluid is more resistive than its dc value suggests.

{\it Conclusion.}\textemdash 
In summary, we have shown how connecting a superconducting sample to a terahertz resonator can lead to a modest but measurable increase in the transition temperature. This effect crucially relies on the presence of the finite-resistance normal fluid, and on the presence of strongly fluctuating pre-formed Cooper pairs. A similar scenario was considered from the perspective of noise magnetometry~\cite{Curtis.2024}, and this raises the interesting possibility of testing these predictions {\it in situ} by utilizing a combination of terahertz resonator control and quantum sensing readout. 

Further extensions of this work to unconventional or other strongly fluctuating superconductors would also be interesting, especially in light of recent experiments~\cite{keren2026cavity,montanaro2026cavity}.
Graphene-based systems may also be a rich opportunity for exploring the interplay of fluctuations~\cite{Curtis.2023} and quantum electrodynamics, and coupling them to terahertz resonators~\cite{helmrich2026cavity,jochl2026absence} may offer a new route for exploring the finite-frequency dynamics of these materials in the presence of strong fluctuations.
Direct confirmation of these theories may be obtainable in quantum simulators based on Josephson-junction arrays, where single-site control and readout could further identify the role that vortices play, and potentially add terahertz resonators as a useful new tool for establishing long-range coherence in these systems at higher temperatures. 
Finally, it would be interesting to apply the Casimir formalism accounting for strong order-parameter fluctuations to multimode geometries in which the coupling to the sample scales extensively \cite{bimonte2005towards}, such as polar substrates~\cite{keren2026cavity} like SrTiO$_3$, whose soft phonon could effectively plays the role of the resonator.

\acknowledgements{We thank Andrea Cavalleri and Sambuddha Chattopadhyay for fruitful discussions. F.L.~acknowledges helpful discussions with Oleg Kotov and Jeremy Munday and was supported by the Quantum Center Research Fellowship and the Dr Alfred and Flora Spälti Fonds. F.L., J.C., and E.D. are supported by the SNSF project 200021 212899, the SNSF Sinergia grant CRSII--222792, NCCR SPIN, a National Centre of Competence in Research (SNSF grant number 225153). J.F.~acknowledges funding from the Swiss National Science Foundation (SNSF) (grant number 10000397), and by the European Research Council (ERC) (grant COLLECTIVE).}

\appendix

\section{Circuit quantization}
\label{app:circuitQED}

In this Appendix, we derive the Euclidean effective action in
Eq.~\eqref{eqn:actioncircuit} of the quantized circuit in Fig.~\ref{fig:schematic}(b).

The loop in Fig.~\ref{fig:schematic}(b) carries a single charge coordinate $Q(\tau)$,
with current $I=\dot Q$. The two-fluid film is internally a parallel combination of
$L_k$ and $R_\mathrm{nf}$, so we introduce a second coordinate $Q_s$, defined as the
charge through the kinetic branch. Consequently, the charge through the resistive
normal-fluid branch is $q_R = Q - Q_s$.

Following the Caldeira--Leggett model for dissipation in quantum systems~\cite{Caldeira.1983rkp, devoret1995quantum}, the resistive branch is
represented by a bath of oscillators $\{x_k\}$ bilinearly coupled to $q_R$. The imaginary-time action of the circuit then reads
\begin{align}
  S_E[Q,Q_s,\{x_k\}] &= S_R[Q] + S_{\rm film}[Q,Q_s,\{x_k\}], \label{eq:quantized_circuit}\\
  S_R[Q] &= \int_0^\beta\! d\tau\, \Big[\tfrac12 L_R\dot Q^2 + \frac{Q^2}{2C_R}\Big], \\
  S_{\rm film}[Q,Q_s,\{x_k\}] &= \int_0^\beta\! d\tau\, \Big[\tfrac12 L_k\dot Q_s^2 + \sum_k\Big(\tfrac12 m_k\dot x_k^2 \nonumber
  \\
  &+ \tfrac12 m_k\omega_k^2 x_k^2 
  - c_k x_k q_R + \frac{c_k^2}{2m_k\omega_k^2}\,q_R^2\Big)\Big].
\end{align}
The spectral density of the bath oscillators is given by
$J(\omega)=\frac{\pi}{2}\sum_k \frac{c_k^2}{m_k\omega_k}\delta(\omega-\omega_k)$.
For a dissipative element with impedance $Z_{\rm diss}$ we choose
$J(\omega)=\omega\,\mathrm{Re}\,Z_{\rm diss}(\omega)$. For the Ohmic case considered in
the main text, $Z_{\rm diss} = R_\mathrm{nf}$ and hence $J(\omega)=R_\mathrm{nf}\,\omega$.

The free energy of the system, $F/T = -\ln\mathcal Z$, follows from the partition function
\begin{equation} \label{eq:partition}
  \mathcal{Z} = \int \mathcal{D}Q\, \mathcal{D}Q_s \prod_k\mathcal{D}x_k \;
  e^{-S_E[Q,Q_s,\{x_k\}]} .
\end{equation}
Note that we denote partition functions by a calligraphic $\mathcal{Z}$, to distinguish them from impedances $Z$. We are interested only in the part of the free energy that arises from the coupling between resonator and film. The free energies of the isolated film (already contained in the Ginzburg--Landau functional) and of the bare resonator (independent
of $n_s$) are removed by defining
\begin{equation} \label{eq:SLC_ratio_def}
  e^{-S_{\rm LC}} \equiv \frac{\mathcal Z}{\mathcal Z_R\, \mathcal Z_{\rm film}} ,
\end{equation}
where $\mathcal Z_R = \int \mathcal{D}Q\, e^{-S_R[Q]}$ is the partition function of the
resonator with the film replaced by a short circuit, and
\begin{align} \label{eq:Zfilm}
\mathcal Z_{\rm film} = \int \mathcal{D}Q_s \prod_k \mathcal{D}x_k\,
e^{-S_{\rm film}[0,Q_s,\{x_k\}]},
\end{align}
that of the film and its bath with open terminals
($Q=0$). All path-integral measures, as well as the partition
function of the free bath, appear identically in numerator and denominator of
Eq.~\eqref{eq:SLC_ratio_def} and cancel.

To evaluate the partition function in Eq.~\eqref{eq:partition}, we expand the charges in Matsubara frequencies,
$Q(\tau)=\beta^{-1/2}\sum_m Q_m e^{-i\nu_m\tau}$ with $\nu_m = 2\pi m T$, $\beta=1/T$
and $Q_{-m}=Q_m^*$, and likewise for $Q_s$ and $q_R$. The resonator action becomes
\begin{equation}
  S_R[Q] = \tfrac12 \sum_m \nu_m Z_R(i\nu_m)\, |Q_m|^2, 
\end{equation}
where the resonator impedance reads $ Z_R(i\nu_m) = \nu_m L_R + 1/(\nu_m C_R)$. The bath oscillators are Gaussian and can be integrated out exactly at fixed $q_R$,
\begin{equation}
  \int \prod_k\mathcal{D}x_k\; e^{-S_E[Q,Q_s,\{x_k\}]} = \mathcal{Z}_B\,
  e^{-S_{\rm eff}[Q,Q_s]} ,
\end{equation}
where $\mathcal{Z}_B$ is the partition function of the decoupled bath and
\begin{align}
  S_{\rm eff}[Q,Q_s] &= S_R[Q] + \tfrac12\sum_m \Big[\nu_m^2 L_k |Q_{s,m}|^2
  + \kappa(\nu_m)\, |q_{R,m}|^2\Big], \label{eq:Seff_QQs}\\
  \kappa(\nu) &= \frac{2}{\pi}\int_0^\infty \! d\omega\,
  \frac{J(\omega)\,\nu^2}{\omega(\nu^2+\omega^2)} \\
  &= |\nu|\, Z_{\rm diss}(i|\nu|).
\end{align}
In the last step, we used that the passive impedance $Z_{\rm diss}$ satisfies the Kramers--Kronig relations. Note that for the Ohmic case, we have $\kappa(\nu) = |\nu| R_\mathrm{nf}$.

The action~\eqref{eq:Seff_QQs} is quadratic in $Q_s$. Completing the square at fixed
$Q_m$ gives
\begin{multline} \label{eq:Seff_intem}
  \int\!\mathcal{D}Q_s \prod_k\mathcal{D}x_k\; e^{-S_E}
  \\= \mathcal Z_{\rm film}\,
  \exp\!\Big\{-\tfrac12\sum_m \nu_m \Big[Z_R(i\nu_m) + Z_{\rm sc}(i\nu_m)\Big]|Q_m|^2\Big\},
\end{multline}
where we identified the prefactor to be $\mathcal Z_{\rm film}$ defined in Eq.~\eqref{eq:Zfilm}. For the Ohmic bath, we find that $Z_{\rm sc}(i\nu_m)$ is given by Eq.~\eqref{eq:Zsquar}, i.e. is the impedance of the parallel combination of the kinetic inductance $L_k = N_\square L_{k\square}$ and the normal fluid resistance $R_\mathrm{nf} = N_\square R_{\mathrm{nf}\square}$ of the superconductor. The
film therefore enters only through its impedance in series with the resonator, i.e. through $Z_R + Z_{\rm sc}$. The remaining Gaussian integral over $Q$ gives
\begin{equation} \label{eq:Z_pre_final}
  \mathcal Z = \mathcal Z_{\rm film}\, \mathcal N_Q \prod_m
  \Big[\nu_m\big(Z_R(i\nu_m) + Z_{\rm sc}(i\nu_m)\big)\Big]^{-1/2},
\end{equation}
while the same integral without the film gives
\begin{align} \label{eq:Zr}
    \mathcal Z_R = \mathcal N_Q \prod_m [\nu_m Z_R(i\nu_m)]^{-1/2}
\end{align}
 with the same normalization $\mathcal N_Q$. Inserting Eqs.~\eqref{eq:Z_pre_final} and \eqref{eq:Zr} into Eq.~\eqref{eq:SLC_ratio_def}, we find 
\begin{equation}
  S_{\rm LC} = \frac12 \sum_{\nu_m} \ln\!\Big[1+\frac{Z_{\rm sc}(i\nu_m)}{Z_R(i\nu_m)}\Big],
\end{equation}
which is Eq.~\eqref{eqn:actioncircuit} of the main text.

\section{Multi-capacitor model}
\label{app:capacitors}

Here, we analyze two different coupling schemes, galvanic and capacitive coupling, and compare both to the model used in the main text, in which the sample impedance is in series with the $LC$ resonator.

If the sample is galvanically coupled to the circuit by, e.g., placing it along the arm of the split ring resonator, its impedance is in series with the resonator, as assumed in the main text. Since the sample is much smaller than the resonator structure, the inductance and capacitance of the resonator are essentially unchanged by its presence.

\begin{figure}
    \centering
    \includegraphics[width=\linewidth]{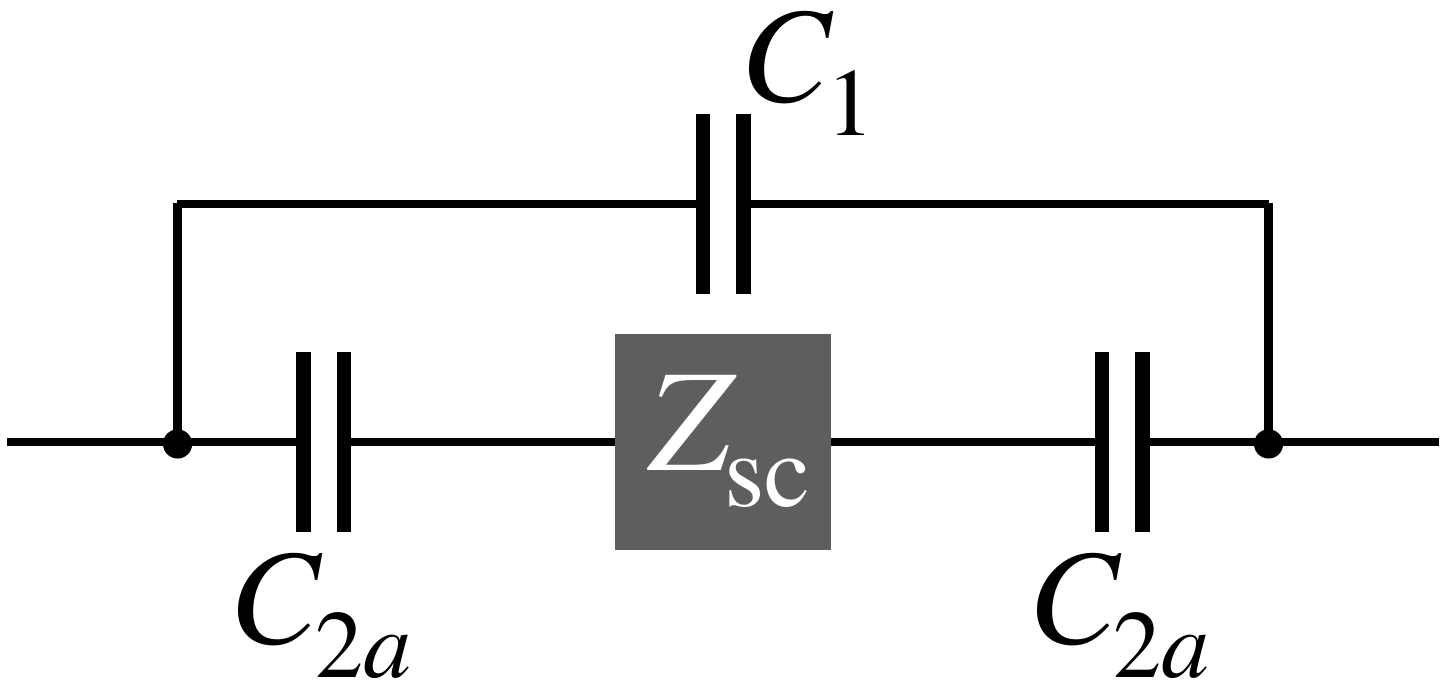}
    \caption{Effective circuit model describing the capacitively coupled sample.}
    \label{fig:capacitances}
\end{figure}

Capacitive coupling instead places the sample inside the capacitive region without galvanic contact, as realised in Ref.~\cite{wang2026evidence} and illustrated in Fig.~\ref{fig:schematic}(a). The impedance of that region is then described by the network of Fig.~\ref{fig:capacitances}: field
lines leaving one arm of the resonator either terminate on the opposite arm, giving an arm-to-arm capacitance $C_1$, or on the sample, giving two arm-to-sample capacitances $C_{2a}$. Our model of an $LC$ resonator in series with the sample corresponds to the limit $C_1 \ll C_{2a}$. This limit is favored by reducing the arm-to-sample distance, increasing the arm-to-arm distance, or placing the sample on a high-index substrate, which draws field lines from the vacuum gap to the substrate on which the sample is placed \cite{wang2026evidence}. In this limit $C_1 \ll C_{2a}$, the sample is again in series with an $LC$ resonator whose inductance remains that of the bare resonator, $L_R$, while its capacitance becomes $C_R = C_{2a}/2$. Note that this means that for capacitive coupling $\Omega_\mathrm{LC} = 1/\sqrt{L_R C_R}$ is no longer the resonance frequency of the empty resonator in our model.

\section{Bare BCS control and Zimmermann model of optical conductivity}
\label{app:bare_BCS_zimmermann}

Here, we first evaluate the resonator-induced bare BCS shift of the critical temperature $T_\mathrm{c}$ excluding any fluctuation effects and using the two-fluid model for the sample response. We then compare the $T_\mathrm{c}$-shift obtained using the two-fluid model with the one obtained from the Zimmermann model, which accounts for a finite impurity scattering rate.

\begin{figure}
    \centering
    \includegraphics[width=\linewidth]{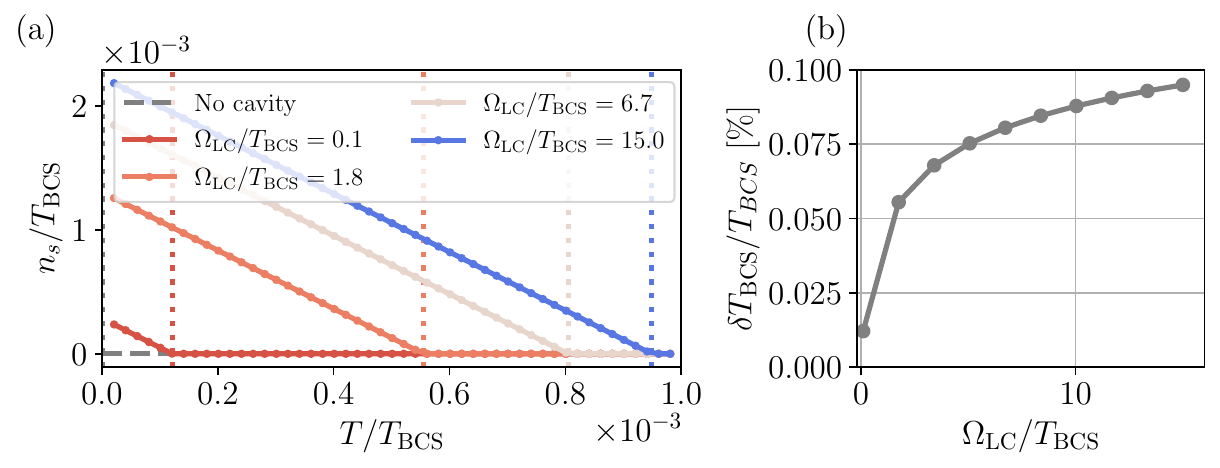}
    \caption{(a) Superfluid density $n_s^\ast$ as a function of $T$ without the cavity (gray dashed line), and with cavities of increasing resonance frequency $\Omega_{\rm LC}$. 
    We take $\sigma_\square = 0.64$ and $L/\xi_{\rm GL}(0) = 100$ and fix the resonator to have $(C_R/e^2)/(e^2 L_R) = 400$ \cite{jeannin2019ultrastrong}. 
    (b) The mean-field critical temperature increases with increasing resonator frequency, reaching $\sim 0.1\%$ at $\Omega_{\rm LC} \sim 15 T_{\rm BCS}$.
    For $T_{\rm BCS} \sim 6.4$K this corresponds to $\Omega_{\rm LC}/2\pi \sim 2 $THz.
    }
    \label{fig:BCS}
\end{figure}

Performing the minimization in Eq.~\eqref{eq:nstae_BCS}, we can find the superfluid density as a function of temperature in the presence of the resonator. The result is shown in Fig.~
\ref{fig:BCS}(a) for different resonator frequencies $\Omega_\mathrm{LC}$ and all other parameters as in Fig.~\ref{fig:BKT} of the main text. We see that a finite superfluid density is also found above the critical temperature in the absence of the resonator $T_\mathrm{BCS}$, showing how the resonator increases the critical temperature. The corresponding shift in the critical temperature as a function of resonator frequency is shown in Fig.~\ref{fig:BCS}(b).

In Fig.~\ref{fig:BCS} and throughout the main text, the optical response of the dirty superconductor is described by the two-fluid model, in which a Drude normal state and a purely inductive condensate channel sit in parallel. This is a phenomenological ansatz, whose microscopic counterpart is the conductivity of Zimmermann \textit{et al.}~\cite{zimmermann1991optical}, which evaluates the BCS current--current response for an arbitrary impurity scattering rate $\gamma$ and reduces to Mattis--Bardeen theory \cite{Mattis.1958} in the dirty limit $\gamma \gg \Delta$. On the imaginary frequency axis, the Zimmermann conductivity can be expressed as \cite{bimonte2010optical}
\begin{align}\label{eq:sigma_split}
    \sigma_\mathrm{Z}(\mathrm{i}\nu) = \frac{\sigma_0}{1 + \nu \tau} + \delta\sigma_{\rm BCS}(\mathrm{i}\nu),
\end{align}
where the first term is the normal metal Drude response with $3$D normal conductivity $\sigma_0 = \sigma_\square/d$, the scattering time $\tau = 1/\gamma$ and the second term is the strictly positive BCS correction:
\begin{align}\label{eq:dsigma_imag}
    \delta\sigma_{\rm BCS}(\mathrm{i}\nu) & = \frac{\sigma_0\gamma}{\nu}
    \int_{-\infty}^{\infty}\!\frac{d\varepsilon}{E}\,
    \tanh\!\Big(\frac{E}{2T}\Big)\, \mathrm{Re}\big[G(\mathrm{i}\nu,\varepsilon)\big], \\
    G(\mathrm{i} \nu,\varepsilon) &= \frac{\varepsilon^2 Q + (Q + \mathrm{i}\gamma)\,A}
        {Q\big[\varepsilon^2 - (Q + \mathrm{i}\gamma)^2\big]}, \\
    A(\mathrm{i} \nu,E) &= E(E + \mathrm{i} \nu) + \Delta^2, \\
    Q(\mathrm{i} \nu,E) &= \sqrt{(E + \mathrm{i} \nu)^2 - \Delta^2},
\end{align}
where $\mathrm{Im}\, Q > 0$ and $E = \sqrt{\varepsilon^2 + \Delta^2}$.

Note that when the Zimmermann and two-fluid conductivities are normalised to the same superfluid weight $\rho_{\rm s} = \lim_{\nu \to 0} \nu\,\sigma(i\nu)$, with $\rho_{\rm s} = 4e^2 n_s $ they differ only in how rapidly $\sigma(i\nu)$ falls off once $\nu \gtrsim 2\Delta$. This means that for frequencies $\nu \ll 2\Delta$, the two fluid model provides a good approximation. 

In order to test this, we evaluate the bare BCS shift with the Zimmermann conductivity instead of the two-fluid one for the parameters used in the main text. For the Zimmermann model, the impedance $Z_\mathrm{Z}$ of the sample is given by $Z_\mathrm{Z} = l/(A\sigma_\mathrm{Z})$, with the cross-sectional area of the superconductor $A = dW$, where $l$ and $W$ are the length and width of the sample and we used $l = W = L$ for all simulations. For high frequencies, $\sigma_Z \to \sigma_0/( \tau \nu) $ such that we find that $L_\infty  \equiv \lim_{\nu_n\to \infty}Z_\mathrm{Z}(\mathrm{i}\nu_n)/\nu_n $ is given by $L_\infty = l \tau /(A\sigma_0)\neq 0 $. We therefore define
\begin{align}
    Z_\mathrm{Z}(\mathrm{i}\nu_n) = Z_{\mathrm{Z},\infty}(\mathrm{i}\nu_n) + Z_{\mathrm{Z,CL}}(\mathrm{i}\nu_n),
\end{align}
with $Z_{\mathrm{Z,CL}} = Z_{\mathrm{Z}}- Z_{\mathrm{Z},\infty}$ and $Z_{\mathrm{Z},\infty}(\mathrm{i}\nu_n) = L_\infty \nu_n $.
$Z_{\mathrm{Z,CL}}$ then is positive definite and falls off sufficiently fast for high frequencies such that it can be modeled by a Caldeira-Leggett bath as outlined in Appendix~\ref{app:circuitQED}. As a result, we can still use the quantization scheme and effective action derived in Appendix~\ref{app:circuitQED} with the two-fluid impedance of the sample replaced by $Z_\mathrm{Z}$.

\begin{figure}
    \centering
    \includegraphics[width=0.95\linewidth]{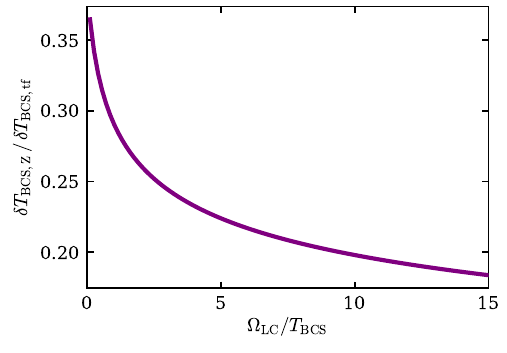}
    \caption{Ratio of the critical temperature shift $\delta T_\mathrm{BCS}$ of the bare BCS model obtained with the Zimmermann ($\delta T_{\rm BCS, Z}$) and the two-fluid ($\delta T_{\rm BCS, tf}$) model, respectively, for the parameters used in the main text and $\gamma = 10^3 T_\mathrm{BCS}$. The result changes by less than $15\,\%$ for any $\gamma \ge 10^2 T_\mathrm{BCS}$. }
    \label{fig:Zimmermann_comp}
\end{figure}

The ratio between the bare BCS shift in the critical temperature within the two-fluid model and within the Zimmermann model is shown in Fig.~\ref{fig:Zimmermann_comp}. As we can see, we find that the two models give a resonator-induced renormalization of $T_\mathrm{BCS}$ that is of the same order of magnitude and the Zimmermann shift is smaller by a factor of about $0.2-0.4$ in the parameter range considered in the main text. Note also that as $\delta \sigma_\mathrm{BCS}\ge 0$ and $Z_Z$ decreases for increasing $\sigma_\mathrm{Z}$, we find that also within the Zimmermann model the resonator always enhances the critical temperature. Thus, also using the Zimmermann model, we find exactly the opposite sign of the critical-temperatures shift to what one obtains when only the superfluid density is considered, in which case an increase in the superfluid density creates spectral weight instead of relocating it from the normal fluid background. 

\section{BKT}
\label{app:BKT}
Here we provide more details about the renormalization group (RG) approach employed here. 
To reach length scales beyond the coherence length $\xi_{\rm GL} \sim 1/\sqrt{1-t}$, we integrate the Kosterlitz RG equations~\cite{Kosterlitz.1974,Jose.1977,Kosterlitz.2016} over scale $b = \log(\ell/\xi_{\rm GL})$. 
The RG flow equations read
\begin{subequations}
\begin{align}
& \frac{dK}{db} = -4 \pi^3 y^2 K^2 , \\
& \frac{dy}{db} = (2 - \pi K) y,  \\
& \frac{dS}{db} = 2 S - 2\pi y^2 ,
\end{align}
\end{subequations}
where $K = n_s/T$ is the dimensionless renormalized phase stiffness, and here we have also included the scaling equation for the action $S(b)$ of a patch size $\mathrm{e}^{2b} \xi_\mathrm{GL}^2$ with corresponding free energy at length scale $b$ given by $F(b) = T S(b)$. 
The initial value of $K$ at $b =0$ ($K_0$) is taken as a free parameter, which we variationally minimize the action over. 
We then model the initial vortex fugacity and action as 
\begin{subequations}
\begin{align}
& y_0 = c\exp\left( - \frac{\pi K_0}{2} \right),  \\
&S_0( K_0 ) = \frac12  \left[  -K_0   + \frac{7 \zeta(3)e^2}{2 \pi^3  \sigma_\square}\frac{t}{1-t} K_0^2  \right].
\end{align}
\end{subequations}
The coefficient $c$ is phenomenologically set at $c \sim 0.1$~\cite{Curtis.2024}, corresponding to a dilute vortex approximation. 
We then integrate the RG flow.

In the algebraically ordered phase this is terminated at $b_f = \log(L/\xi_{\rm GL})$ but in the normal state, $y$ diverges as $b \to \infty$ and so we should in fact terminate the scaling at $b_{\rm max} = \min( \log L/\xi_{\rm GL} , \log \xi_+/\xi_{\rm GL})$ where $\xi_+$ is the length scale at which $y = 1$. 
From there we then use the naive scaling for an uncorrelated system; namely, 
\begin{equation}
S(K; \ell = L) = (L/\xi_+)^2 S(K; \ell =\xi_+).
\end{equation}
We also must use the scaling for the screened phase stiffness, which we will assume to be in this case again 
\begin{equation}
K(L) = (\xi_+/L)^2 K(\xi_+),
\end{equation}
which roughly follows from a Thomas-Fermi screening ansatz by free vortices at scales beyond $\xi_+$~\cite{Halperin.1979}. 

This procedure then yields two quantities.
The first is the renormalized action at the scale of the sample $S_R(K)$, taking into account the energy due to vortices, which is neglected by simply scaling the bare BCS action by the number of patches.
The second quantity this yields is the renormalized phase stiffness $n_s^{(R)} = T K$. 
This can be used, in a two-fluid model, to compute the kinetic inductance of the sample when it is incorporated into the circuit as 
\begin{equation}
L_{k\square} = \frac{1}{n_s^{(R)}}\frac{1}{(2e)^2}  = \frac{1}{4 e^2 T K}.
\end{equation}

In the main text, we have taken as the variational functional the bare action $S_0$, expressed as a functional of the renormalized superfluid density. 
This is equivalent to truncating the scaling equation for $S$ at order $O(1)$, so that $dS/db = 2S + O(y^2)$.
Doing so, we find the standard textbook curve for superfluid density vs. temperature, which has a sharp but continuous drop near $T_{\rm BKT}$.
In reality however, it is not clear that this is conceptually the correct procedure, and instead one should perhaps minimize the renormalized action $S_R$, obtained by including the free-energy of vortices, i.e. including the contribution of $-2\pi y^2$.

\begin{figure}
    \centering
    \includegraphics[width=\linewidth]{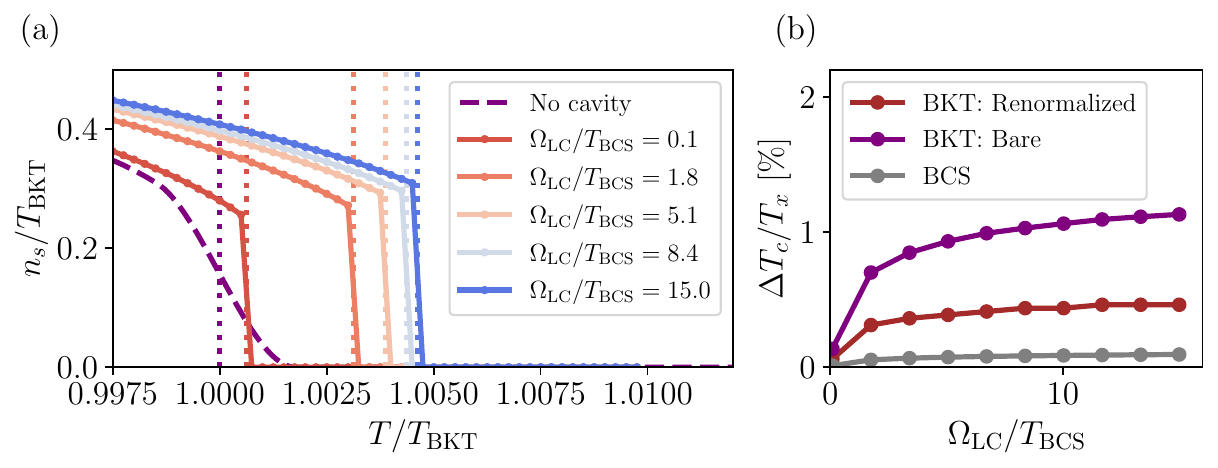}
    \caption{(a) Superfluid density vs. temperature without resonator, and in presence of resonator for $\sigma_\square = 0.64$ and $L/\xi_{\rm GL}(0) = 100$.
    We see the resonator tends to increase the transition temperature, but also seems to drive the transition first order, with a substantial jump appearing in $n_s(T)$ even for small $\Omega_{\rm LC}$. 
    (b) Comparison of enhancement for different models of the superconductor, still with $(C_R/e^2)/(e^2L_R) = 400 $.
    We see the two different BKT modelings result in a qualitatively similar form, though differ by a factor of roughly two; both are substantially more pronounced than the BCS model.}
    \label{fig:BKT_renormalized}
\end{figure}

We have also computed the results using the renormalized action $S_R$ and find qualitatively similar results, shown in Fig.~\ref{fig:BKT_renormalized}. 
However, numerically this tends to lead to a very sharp jump in the superfluid density vs. temperature even at finite length-scales $b$.
This perhaps indicates that once the free-energy of vortices is taken into account, the transition is driven first-order, or it could signal a rapid breakdown of the perturbative scaling equations.
In either case, we include the results here for completeness. 
However, to our knowledge this distinction is not discussed in the literature and is in fact an open question regardless of the coupling to a resonator or not.

\end{document}